\documentclass[cernpreprint, atlasdraft=true, texlive=2016, UKenglish, texmf]{atlasdoc}
 
\usepackage{atlaspackage}
\usepackage[backref=false, block=none]{atlasbiblatex}
 
\usepackage{atlasphysics}

\graphicspath{{logos/}{figures/}}
 
\usepackage{ANA-EXOT-2019-36-PAPER-defs}
\usepackage{PMGRefs-defs}
\usepackage{makecell}
\usepackage{cancel}

\AtlasTitle{Search for new phenomena in three- or four-lepton events in $pp$ collisions at $\sqrt{s} = 13$~TeV with the ATLAS detector}
 
\AtlasAbstract{
A search with minimal model dependence for physics beyond the Standard Model in events featuring three or four charged leptons ($3\ell$ and $4\ell$, $\ell = e,\mu$) is presented. The analysis aims to be sensitive to a wide range of potential new-physics theories simultaneously. This analysis uses data from $pp$ collisions delivered by the Large Hadron Collider at a centre-of-mass energy of $\sqrt{s} = 13$~TeV and recorded with the ATLAS detector, corresponding to the full Run 2 dataset of 139~fb$^{-1}$. The $3\ell$ and $4\ell$ phase space is divided into 22 event categories according to the number of leptons in the event, the missing transverse momentum, the invariant mass of the leptons, and the presence of leptons originating from a $Z$-boson candidate. These event categories are analysed independently for the presence of deviations from the Standard Model. No statistically significant deviations from the Standard Model predictions are observed. Upper limits for all signal regions are reported in terms of the visible cross-section.
}
 
\AtlasJournalRef{Phys.\ Lett.\ B 834 (2022) 136832}
\AtlasDOI{10.1016/j.physletb.2021.136832}
 
\author{The ATLAS Collaboration}
 
\AtlasRefCode{EXOT-2019-36}

\PreprintIdNumber{CERN-EP-2021-063}
\hypersetup{pdftitle={ATLAS document},pdfauthor={The ATLAS Collaboration}}
 
\begin{document}
 
\maketitle

\section{Introduction}

Despite the success of the Standard Model (SM)~\cite{GLASHOW1961579,PhysRevLett.19.1264,HOOFT1971167,THOOFT1972189} in describing the interactions of elementary particles, there remain observations that suggest the existence of additional phenomena~\cite{dark_matter,dark_energy,NO1,NO2,Sakharov_1991}. Many theories of physics beyond the Standard Model (BSM theories) have been proposed that feature final states in high-energy proton--proton ($pp$) collisions with exactly three or four leptons (\tl and \fl, where $\ell = e,\mu$ in this paper). An example is supersymmetry (SUSY), where neutralino and chargino production~\cite{SUSY-2018-06} yields three leptons and a neutrino through an intermediate $WZ$ state, and where di-Higgs production~\cite{huang2019dihiggs} yields four or more leptons. Furthermore, enhanced flavour-changing decay cross-section of top quarks at the loop level may lead to anomalous production of \tl final states with respect to the SM~\cite{afik2021new}. Multiple types of seesaw models can produce multilepton final states alongside neutrinos~\cite{del_Aguila_2009,Ma_1998,Gu_2006,Franceschini_2008,t3ss}. Adding an additional Higgs triplet to the SM Lagrangian~\cite{GEORGI1985463,PhysRevD.44.837} potentially leads to a doubly charged Higgs particle~\cite{Dev_2018,EXOT-2016-07} which can decay into two leptons, leading to a four-lepton final state if produced in pairs. Theories predicting such a particle include left--right symmetric models~\cite{PhysRevD.40.1546,Setzer_2007}, scalar singlet dark matter~\cite{Bhattacharya_2017} and the Zee--Babu model~\cite{PhysRevD.77.093013}. Conclusive evidence for any of these BSM theories has thus far been elusive. Many dedicated analyses within the LHC experimental collaborations are being performed to search for such evidence. However, the vast number of theories means that it would be difficult to perform a dedicated analysis for each model. This motivates the desire to establish instead a search that does not rely on a specific model for its signal description (henceforth called `model-independent'), which can cover a wide range of signatures to seek indicators of exotic physics.
 
The analysis presented here is committed to investigating a large phase space while making few prior assumptions about the nature of new-physics processes.
As such, it is expected to be sensitive to a large number of signals that could be populating the ATLAS data, albeit with a lower sensitivity than a dedicated analysis could achieve.
Partial overlap with dedicated analyses is expected, but since these are typically tuned to specific models they do not consider the full phase space.
For instance, Ref.~\cite{SUSY-2018-02} also studies \fl final states, but does not consider events with low four-lepton invariant masses and missing transverse momenta.
The present analysis does not attempt to exploit very distinctive features, such as e.g. resonances in invariant-mass distributions, and instead entails a substantially more inclusive selection than is typical of searches for specific BSM theories, Nevertheless, a comparison with a few benchmark models indicates that its sensitivity is not greatly reduced compared to that of more dedicated searches.
 
The analysis aims to uncover evidence of BSM physics. Failing that, it can provide a set of upper limits on the visible cross-section, which can be reinterpreted as upper limits on BSM models of interest. The upper limits on two benchmark models derived in this fashion, a Type-III seesaw model~\cite{t3ss0} and a doubly-charged Higgs model~\cite{Huitu:1996su} are compared with those obtained using dedicated analyses.
 
For this analysis, the full $pp$ dataset collected by the ATLAS experiment during the 2015--2018 data-taking period is used, corresponding to an integrated luminosity of 139~\ifb\  delivered by the LHC.
Events featuring exactly three or four charged leptons are categorised into signal regions based on the invariant mass of the leptons, the missing transverse momentum and the presence of a lepton pair compatible with originating from a $Z$-boson decay.
The observations in such regions are individually used to probe for the presence of a BSM signal.
Control regions are established to extract a normalisation of the most prominent SM backgrounds, which are leptonically decaying $WZ$ and $ZZ$ diboson pairs, and for the estimation of the contribution from interactions producing lepton candidates from heavy-flavor hadrons decays or hadronic particles misidentified as leptons.
In each region, a SM-only hypothesis is compared with a hypothesis assuming the SM plus an additional number of BSM events as a free parameter.
 
A previous general multilepton search with the ATLAS detector was performed using 20.3~\ifb\ of $pp$ collisions at $\sqrt{s} = 8$~\TeV~\cite{EXOT-2012-20}. Compared to that search, the current analysis uses a larger dataset collected at \com and assigns \fl events to multiple separate regions, but does not consider hadronic $\tau$-lepton decays. A strategy for a general search was outlined in Ref.~\cite{EXOT-2016-38} using 3.2~\ifb\ of data at \com. That search uses fewer data events and a coarser background estimation, but offers a broader selection of final states, including multilepton final states, and performs tests with additional variables. Furthermore, a similar search, also testing multilepton final states, has been performed by the CMS Collaboration with 137~\ifb\ of $pp$ collisions at \com~\cite{cms_multilep}.
 
\section{ATLAS detector}
 
\newcommand{\AtlasCoordFootnote}{
ATLAS uses a right-handed coordinate system with its origin at the nominal interaction point (IP)
in the centre of the detector and the \(z\)-axis along the beam pipe.
The \(x\)-axis points from the IP to the centre of the LHC ring,
and the \(y\)-axis points upwards.
Cylindrical coordinates \((r,\phi)\) are used in the transverse plane,
\(\phi\) being the azimuthal angle around the \(z\)-axis.
The pseudorapidity is defined in terms of the polar angle \(\theta\) as \(\eta = -\ln \tan(\theta/2)\). Angular distance is measured in units of \(\Delta R \equiv \sqrt{(\Delta\eta)^{2} + (\Delta\phi)^{2}}\).}
 
The ATLAS experiment~\cite{PERF-2007-01} at the LHC is a multipurpose particle detector
with a forward--backward symmetric cylindrical geometry and a near \(4\pi\) coverage in
solid angle.\footnote{\AtlasCoordFootnote}
It consists of an inner tracking detector (ID) surrounded by a thin superconducting solenoid
providing a \SI{2}{\tesla} axial magnetic field, electromagnetic and hadron calorimeters, and a muon spectrometer.
The inner tracking detector covers the pseudorapidity range \(|\eta| < 2.5\).
It consists of silicon pixel (with the insertable B-layer installed before Run~2~\cite{ATLAS-TDR-19,PIX-2018-001}), silicon microstrip, and transition radiation tracking detectors.
In the range $|\eta| < 3.2$, hermetic lead/liquid-argon (LAr) sampling calorimeters provide electromagnetic (EM) energy measurements
with high granularity. The central region, $|\eta| < 1.8$, is additionally instrumented with a thin LAr presampling detector
to correct for energy losses in the inactive material in front of the detector.
A steel/scintillator-tile hadronic calorimeter covers the central pseudorapidity range $|\eta| < 1.7$.
The endcap and forward regions are instrumented with LAr calorimeters
for both the EM and hadronic energy measurements up to \(|\eta| = 4.9\).
The muon spectrometer surrounds the calorimeters. It consists of
three large superconducting air-core toroidal magnets with eight coils each.
The field integral of the toroids ranges between \num{2.0} and \SI{6.0}{\tesla\metre}
across most of the detector.
The muon spectrometer includes three layers of precision tracking
chambers, allowing precise muon momentum measurements up to $|\eta| = 2.7$, and fast detectors for triggering up to $|\eta| = 2.4$.
A two-level trigger system~\cite{TRIG-2019-04} is used to select events.
The first-level trigger is implemented in hardware and uses a subset of the detector information
to accept events at a rate below \SI{100}{\kHz}.
This is followed by a software-based trigger that
reduces the accepted event rate to \SI{1}{\kHz} on average
depending on the data-taking conditions. 
\section{Data and simulated events}
 
This analysis uses $pp$ collision data at a centre-of-mass energy of \com, collected by the ATLAS detector and corresponding to a total integrated luminosity of 139~\ifb. Only data recorded during stable beam conditions with all ATLAS detector subsystems operational~\cite{DAPR-2018-01} have been included. Dilepton triggers~\cite{TRIG-2018-05,TRIG-2018-01} covering all lepton flavour combinations ($ee$, $e\mu$ and $\mu\mu$) are used. The transverse momentum (\pt) requirements of these triggers depend on the data-taking period. For the dielectron trigger, these requirements are 12~\GeV{} in 2015, 17~\GeV{} in 2016, and 24~\GeV{} in 2017--18. For the dimuon trigger, they are 18 and 8~\GeV{} in 2015 and 22 and 8~\GeV{} in 2016--18.\footnote{At least one muon must pass the higher \pt requirement and a second muon at least the lower \pt requirement.} For the mixed-flavour trigger, the \pt requirement is 17~\GeV{} for the electron and 14~\GeV{} for the muon. These trigger choices correspond to the dilepton triggers with the lowest \pt requirements available during each data-taking year.
 
Expected event rates due to SM processes that can result in \tl and \fl final states were estimated using a combination of Monte Carlo (MC) event generation and data-driven techniques.
Event generators based on MC methods were used to estimate the total expected contributions from SM processes producing only prompt leptons.\footnote{A lepton is prompt if none of the particles in its production chain, traced back to the interaction point of the $pp$ collision, are hadrons.} 
The dominant SM backgrounds are the production of two vector bosons decaying leptonically: $WZ$ for the \tl final states and $ZZ$ for the \fl final states. Subleading prompt-lepton backgrounds that contribute are triboson production, and processes which include at least one top quark: $t\bar{t}X$ ($X = W,Z,H$), $tZ$, $t\bar{t}WW$ and $t\bar{t}t\bar{t}$.
 
All diboson and triboson ($VV$ and $VVV$, where $V=W,Z$) production, including off-shell production, was simulated with the \sherpa~2.2.2~\cite{Bothmann:2019yzt} generator. 
The \nnpdfnnlo set of PDFs was used~\cite{Ball:2014uwa}, along with the dedicated set of tuned parton-shower parameters developed by the \sherpa authors.
The matrix element calculations were matched and merged with the \sherpa parton shower~\cite{Schumann:2007mg} based on Catani--Seymour dipole factorisation~\cite{Winter:2003tt,Gleisberg:2008fv} using the MEPS@NLO prescription~\cite{Hoeche:2011fd,Hoeche:2012yf,Catani:2001cc,Hoeche:2009rj}.
Diboson events were generated at next-to-leading-order (NLO) accuracy in QCD for up to one additional parton and at leading-order (LO) accuracy for two and three additional parton emissions.
Electroweak $VVjj$ ($j =$ jet) events were generated at LO. This contribution includes Higgs boson production through vector-boson fusion (with $H\to ZZ$). It also includes triboson processes where one boson decays hadronically, including $VH\to VVV \to VVjj$.
Loop-induced production of $ZZ$ events via gluon--gluon fusion was simulated using matrix elements accurate at LO for up to one additional parton emission for both the fully leptonic and semileptonic final states. This contribution includes $gg\to H \to ZZ$.
 
Triboson events were generated at NLO for the inclusive process and at LO for up to two additional parton emissions.
The virtual QCD corrections were provided by the \openloops\ library~\cite{Buccioni:2019sur,Cascioli:2011va,Denner:2016kdg}. This process includes only on-shell fully leptonic decays.
 
The production of \ttV, \ttWW, \tZq\ and \tttt events was modelled using the \mgamc~v2.3~\cite{Alwall:2014hca} (v2.2 for \ttWW) generator, while \ttH\ events~\cite{HIGG-2018-13} were modelled using the \powhegbox~\cite{Frixione:2007nw,Nason:2004rx,Frixione:2007vw,Alioli:2010xd}~v2 generator with the \hdamp\ parameter set to 1.5~\mtop~\cite{ATL-PHYS-PUB-2016-020}. Events were generated at NLO for \ttV and \ttH. The \tZq, \ttWW and \tttt processes were modelled at LO with their cross-sections normalised to NLO predictions~\cite{Alwall:2014hca}. The \nnpdfnlo~\cite{Ball:2014uwa} PDF was used. The events were interfaced to \PYTHIA{8.210} (8.230 for \tZq)~\cite{Sjostrand:2014zea} using a set of tuned parameters called the A14 tune~\cite{ATL-PHYS-PUB-2014-021} and the \nnpdftwo~\cite{Ball:2014uwa} PDF set.  Collectively, these events are referred to as the top-quark background.
 
Additional processes involving Higgs boson production, apart from those mentioned above, have been found to yield negligible contributions and are not considered explicitly
 
For model-specific interpretation of the analysis, signal samples were generated for Type-III seesaw model heavy leptons and doubly charged Higgs (\dch) particles decaying leptonically. The simplified Type-III seesaw model was included in the \mgamc~v2.3.3 generator at LO with an implementation using FeynRules~\cite{Alloul:2013bka}, described in Ref.~\cite{t3ss0}.  Implementation details are described in Ref.~\cite{t3ss}. Samples of \dch events were generated at LO using the left--right-symmetry package of \PYTHIA{8.186}, which provides the $H^{\pm\pm}$ scenario described in Ref.~\cite{Huitu:1996su}. Implementation details are described in Ref.~\cite{EXOT-2016-07}. The parton shower was provided by \PYTHIA{8.230} (\PYTHIA{8.186}) for the Type-III seesaw (\dch) model using the \nnpdftwo PDF set and the A14 tune~\cite{ATL-PHYS-PUB-2014-021}. Cross-sections for both samples were normalised to NLO.
 
The ATLAS detector simulation~\cite{SOFT-2010-01} employing the $\GEANT$~\cite{Agostinelli:2002hh} framework was used to model the detector response in MC events.
The effect of \pileup was incorporated into the simulation by overlaying additional inelastic $pp$ events onto hard-scatter events.
These were generated with \PYTHIA{8}~\cite{Sjostrand:2014zea} using the A3 tune~\cite{ATL-PHYS-PUB-2016-017} and the \texttt{MSTW2008LO}~\cite{Martin:2009iq} PDF set.
Events were simulated with discrete values for the expected mean number of interactions and then weighted to match the distribution that is observed per bunch crossing in data.
\section{Object selection}
 
All events in the analysis are required to have a primary vertex, defined as the vertex with the highest value of $\sum \pt^{2}$ of its associated tracks, which must include at least two with $\pt > 0.5$~\GeV. Electron and muon candidates are required to originate from the primary vertex.
 
Requirements common to both electron and muon candidates are $\pt > 25$~\GeV{} and $|\eta| < 2.47$.  For the purposes of matching tracks to the primary vertex, the track impact parameters\footnote{The $d_0$ is the transverse impact parameter: the distance of the track from the beam line at the position of closest approach (PCA) in the plane perpendicular to the beam line, while $\sigma(d_0)$ is its uncertainty. The $z_0$ is the $z$ coordinate of the PCA where \dzero is measured, relative to that of the primary vertex.} $d_{0}$ and $z_{0}$ must satisfy $\dzero < 5\,(3)$ for electrons (muons) and $|z_{0} \sin{\theta}| <0.5$~mm. Furthermore, electron and muon candidates are subjected to identification criteria, for which multiple working points are provided~\cite{EGAM-2018-01,MUON-2018-03}. The identification is performed using quality cuts where each working point offers a different trade-off between the rate of false positives and false negatives delivered by the algorithm.
 
Electron candidates are reconstructed using energy clusters measured in the EM calorimeter matched to reconstructed tracks~\cite{EGAM-2018-01}. The identification for the nominal selection in this analysis is based on a combination of detailed tracking and calorimeter information combined into a likelihood discriminant. The \texttt{Tight}~\cite{EGAM-2018-01} working point is used. The range $1.37 < |\eta| < 1.52$ has a significant amount of non-sensitive material in front of the calorimeter, and is therefore excluded.
 
Muon candidates are reconstructed by combining measurements in the ID and the muon spectrometer~\cite{Cornelissen:2007vba}. For this analysis, the \texttt{Medium} identification working point is used for most muon candidates~\cite{MUON-2018-03}, which requires an ID track matched with multiple muon spectrometer precision hits. For muon candidates with $\pt > 300$~\GeV{}, the \texttt{High-\pt} working point~\cite{MUON-2018-03} is used, which places a tighter requirement on the number of muon spectrometer hits, ensuring optimal momentum resolution for highly energetic muons.
 
Both the electron and muon candidates are required to be isolated in the ID. To determine isolation, a cone is placed around the object's track, with an opening angle which is the smaller of $\Delta R = 0.2$ (0.3) for electrons (muons) and $10\,\GeV/ p_{\mathrm{T},\ell}$. The scalar \pt sum of all tracks (excluding the lepton itself) within this cone, $I_R$, must satisfy $I_R / p_{\mathrm{T},\ell} < 0.06$. For muon candidates with $\pt > 50$~\GeV, the opening angle of the isolation cone is always $\Delta R = 0.2$. Electron candidates must also pass a calorimeter-based isolation requirement of $I_R / p_{\mathrm{T},\ell} < 0.06$, this time taking the sum of calorimeter energy deposits as $I_R$, within a cone of $\Delta R = 0.2$.
 
The constituents for jet reconstruction are identified by combining measurements from both the ID and the calorimeter using a particle-flow algorithm~\cite{PERF-2015-09}. Jet candidates are reconstructed from these particle-flow objects using the anti-$k_t$ algorithm~\cite{Cacciari:2008gp,Fastjet} with a radius parameter $R=0.4$. The jet energy scale (JES) and resolution (JER)~\cite{JETM-2018-05} are corrected to particle level using MC simulation. Jets are furthermore required to have $\pt > 20$ \GeV{} and $|\eta| < 2.5$. The jet vertex tagger (JVT)~\cite{PERF-2014-03} is used to test jets that have $\pt<60$~\GeV{} and $|\eta|<2.4$ to suppress those originating from \pileup. 
 
Objects found to have very collinear tracks are considered to be overlapping. Overlaps are resolved through a sequence of rules. This procedure prevents double-counting of particles interacting with different parts of the detector, and provides an optimal classification of these particles. If a muon candidate is found to have a shared ID track with an electron candidate, the electron candidate is rejected. If two electron candidates have shared ID tracks, the one with the lower \pt is rejected. Jets are rejected if they are within $\Delta R^{\prime} = 0.2$ (for overlap removal, the pseudorapidity in $\Delta R$ is substituted with the rapidity, defined as $y=-\ln\frac{E+p_z}{E-p_z}$) of a lepton candidate, except if the candidate is a muon and three or more collinear tracks are found. Subsequently, lepton candidates that are within $\Delta R^{\prime} = 0.4$ of any remaining jets are removed.
 
The missing transverse momentum (\met)~\cite{PERF-2016-07} in a given reconstructed event is computed as a combination of a hard term, the magnitude of the negative vector sum of the \pt of all reconstructed leptons and jets, and a soft term, computed from the momenta of inner-detector tracks that are not matched to any of the selected objects but do originate from the primary vertex.
\section{Analysis strategy}
 
Selected events are separated into different categories, referred to as regions, to maximise the sensitivity to a relatively broad range of potential new phenomena. Signal regions (SRs) are defined as regions to be probed for the presence of such signatures. Criteria that separate these SRs are the number of leptons, the \met and the presence of an on-$Z$ lepton pair: a same-flavour and oppositely charged (SFOC) lepton pair with a dilepton mass within 10~\GeV{} of the $Z$-boson mass of 91.2~\GeV{}.\footnote{A lepton that does not form an on-$Z$ lepton pair with any other lepton in the event is called off-$Z$.} Control regions (CRs) are defined so as to be dominated by particular SM processes which have been well-studied in the ATLAS experiment~\cite{STDM-2018-03,STDM-2018-30}. The full list of regions with their selection criteria, apart from the splitting into different invariant-mass ranges, is shown in Table~\ref{tab:region_table}. All regions are orthogonal to each other, so no event is assigned to more than one region.
The CRs are used to extract the normalisation of the main background processes from data and to constrain the size of the systematic uncertainties of the analysis. Validation regions (VRs) are used to confirm that the predictions for SM background processes are well-modelled.
 
\begin{table}[bt] \begin{center}
\caption{Overview of the regions defined for this analysis. The - symbols indicate that no requirements are made on the variable for that particular region. Additional requirements on SRs, described in the \textit{Other} column, veto events that are used in the CR/VRs from entering into the SRs. The $Z$-pairs column denotes the number of non-overlapping lepton pairs that are same-flavour and oppositely charged and have a dilepton invariant mass within 10 \GeV{} of the $Z$-boson mass of 91.2 \GeV. The off-flavour $\ell$ is the lepton in the \tl event that has a different flavour from the other two leptons; cuts requiring an off-flavour $\ell$ are not applied if all three leptons are of the same flavour.}
\resizebox{\textwidth}{!}{
\begin{tabular}{lllll}
\toprule
\textbf{Region} & \textbf{Particles} & \textbf{\met} & $\boldmath{Z}$\textbf{-pairs} & \textbf{Other}\\
\midrule
\multicolumn{5}{c}{\textbf{ Signal regions } }\\
\midrule
\tl & $3\ell$ & < 50 \GeV & 1 & \textbf{veto} event if $m_{\mathrm{T}}<80$ \GeV{} for off-$Z$ $\ell$ \\
& $3\ell$ & > 50 \GeV & 1 & \textbf{veto} event if $m_{\mathrm{T}}<80$ \GeV{} for off-$Z$ $\ell$ \\
& $3\ell$ & < 50 \GeV & 0 & \textbf{veto} event if $m_{\mathrm{T}}<40$ \GeV{} for off-flavour $\ell$\\
& $3\ell$ & > 50 \GeV & 0 & \textbf{veto} event if $m_{\mathrm{T}}<40$ \GeV{} for off-flavour $\ell$\\
\multicolumn{5}{c}{\textit{ \tl SRs are divided into \minv{} ranges of \emph{0--200}, \emph{200--400}, \emph{400--600} and \emph{>600 \GeV{}}. } }\\
\hline
\fl   & $4\ell$ & < 50 \GeV & 1 & - \\
& $4\ell$ & > 50 \GeV & 1 & - \\
& $4\ell$ & - & 0 & - \\
\multicolumn{5}{c}{\textit{ \fl SRs are divided in \emph{\minv{}} ranges of \emph{0-400} and \emph{>400 \GeV{}}. } }\\
\midrule
\multicolumn{5}{c}{\textbf{ Fake-Factor Estimation Regions } }\\
\midrule
$e$-fakes & $1e$ & $<25$ GeV & - & jets $\geq 1$ \\
$\mu$-fakes & $1\mu$ & $<40$ GeV & - & jets $\geq 2$, at least 1 jet $\pt > 35$ GeV, $|\Delta\phi(\mu,j)| > 2.7$ \\
\midrule
\multicolumn{5}{c}{\textbf{ Validation regions } }\\
\midrule
$3\ell$ On-$Z$ & $ee\mu+e\mu\mu$ & - & 1 &  off-flavour $\ell$: $m_{\mathrm{T}} < 40$ \GeV \\
$3\ell$ Off-$Z$ & $ee\mu+e\mu\mu$ & - & 0 &  2 SFOC leptons. off-flavour $\ell$: $m_{\mathrm{T}} < 40$ \GeV \\
\midrule
\multicolumn{5}{c}{\textbf{ Control regions } }\\
\midrule
$WZ$ & $3\ell$ & - & 1 & off-$Z$ $\ell$: $40 < m_{\mathrm{T}} < 80$ \GeV \\
$ZZ$ & $4\ell$ & - & 2 & \\
\bottomrule
\end{tabular}
}
\label{tab:region_table}
\end{center} \end{table}

A large group of BSM models predict the existence of at least one additional heavy lepton beyond the SM (e.g.\ Ref.~\cite{t3ss0}), either charged or neutral. Such theories often feature final states with one or more neutrinos, which due to being invisible to the detector translates into a non-zero \met. This motivates a selection of SRs separated by a \met cut. A threshold of $\met = 50$~\GeV{} was chosen, which splits the phase space into regions where the \met originates mostly from detector resolution effects and regions where the \met is likely to be due to objects invisible to the detector.
 
SRs are also categorised according to the presence or absence of at least one on-$Z$ lepton pair, and are called on-$Z$ and off-$Z$ SRs respectively. No selection based on charge and flavour is made for these SRs beyond the SFOC pair needed for the on-$Z$ region. Certain heavy BSM particles are expected to decay into lepton pairs without first decaying to an intermediate $Z$-boson (e.g.\ Ref.~\cite{Huitu:1996su}). Off-$Z$ SRs are expected to be sensitive to such signals while excluding the main prompt-lepton background contribution (leptonically decaying $WZ$ and $ZZ$ vector-boson pairs, which are likely to be on-$Z$). Furthermore, on-$Z$ SRs include the few events where three leptons combine into multiple valid on-$Z$ pairs.
 
SRs are further split according to the distribution of the invariant mass (\minv) of all leptons in the event.
Four divisions are established to construct the \tl regions: 0--200~\GeV, 200--400~\GeV, 400--600~\GeV, and ${>}600$~\GeV. Two divisions are established for the \fl regions: 0--400~\GeV{} and ${>}400$~\GeV. This leads to 22 SRs in total.
 
Two CRs are defined: a \tl CR for the $WZ$ background and a \fl CR for the $ZZ$ background. The $WZ$ control region requires an on-$Z$ lepton pair and a third off-$Z$ lepton which has a transverse mass ($m_{\mathrm{T}}$, defined as $m_{\mathrm{T}} = [2 \pt^\ell \met (1 - \cos(\Delta\phi(\ell,\met)))]^{1/2}$), of $40 < m_{\mathrm{T}} < 80$ \GeV{}, which captures leptons originating from a $W$-boson decay. The $ZZ$ control region requires four leptons to form two on-$Z$ lepton pairs. The SRs are separated from the $WZ$ CR through their flavour composition or by requiring $m_{\mathrm{T}} > 80~\GeV$, and from the $ZZ$ CR by vetoing events with 4 on-$Z$ leptons. Both CRs are used to constrain the two normalisation factors of their corresponding backgrounds. These normalisation factors are free parameters in the statistical analysis.
 
A data-driven technique is used to estimate backgrounds with at least one fake lepton, referred to as the fake-lepton background, in the SRs, CRs and VRs. Fake leptons are either non-prompt leptons or hadrons misidentified as leptons by the detector. The primary sources of such events are the $Z+\text{jets}$ and \ttbar processes which have two prompt leptons and at least one fake lepton. The yield of fake-lepton background events is measured separately for electrons and muons using the fake-factor method, which is described in Ref.~\cite{EXOT-2012-20}. Dedicated regions containing a single lepton candidate are established using data collected by the single-lepton triggers.
Selection requirements for these regions are imposed to ensure a large number of events with fake leptons, in order to reduce the statistical uncertainty of this contribution.
Requirements are based on the \met of the event ($\met <25~\GeV$ for electrons, $\met <40~\GeV$ for muons) and on the number of jets in the event ($\geq 1$ for the electron region, $\geq 2$ for the muon region). For the muon, there must also be at least one jet ($j$) with $\pt > 35~\GeV$ and $\Delta\phi(\mu,j) > 2.7$, called the tag jet.
For each region, an adjacent `anti-ID' region is established, with orthogonal selection criteria of the identification and isolation algorithms.
The anti-ID regions admit non-isolated electrons and muons, as well as isolated but \texttt{Loose}~\cite{EGAM-2018-01} electrons, but veto events which satisfy the nominal selection criteria. The anti-ID selection is tuned for a large number of fake leptons, while the stringent working points of the nominal selection suppress these fake leptons, to ensure the analysis is robust against fake-lepton contamination.
A ratio is computed from the event rates of these two regions. This ratio is called the fake factor and is parameterised as a function of the \pt and $\eta$ of the lepton.
An anti-ID region is also established for each signal, control and validation region of the analysis. These anti-ID regions use the same selection criteria, except that one or more of their leptons passes the alternative identification and isolation requirements. Using the fake factor, the yield of fake-lepton background events is extrapolated from each anti-ID region to its nominal counterpart. Prompt backgrounds, estimated through MC methods and normalised to their theoretical cross sections, are subtracted from observed data in the fake-factor estimation regions prior to calculation of the fake factor, and in the anti-ID regions prior to extrapolation.
 
Certain subselections of $ee\mu+e\mu\mu$ events are designated as on-$Z$ and off-$Z$ VRs. These are used to check that the computed fake factors transfer correctly from the regions where they are calculated to the regions in which they are applied. The on-$Z$ and off-$Z$ VRs consist of different ratios of fake-lepton sources: the on-$Z$ VR is more sensitive to $Z$+jets events than the off-$Z$ region, while the reverse is true for \ttbar events. Furthermore, both VRs have a substantial $WZ$ contribution which can be validated. In the off-$Z$ VR the proper modelling of the $WZ$ MC simulation of off-shell $Z$ decays is confirmed. For this VR, a SFOC pair of off-$Z$ leptons is required.
Both validation regions target, through a $m_\mathrm{T}$ requirement of $m_{\mathrm{T}}(\ell,\met) < 40~\GeV$, a third lepton that is likely to be fake. Only mixed-flavour final states are selected for these VRs so that the choice of the third lepton, assumed to be the fake lepton (or the lepton due to $W$-boson decay), is unambiguous.

\section{Systematic uncertainties}
 
Systematic uncertainties affect the precision of the predicted background contributions. Two classes of systematic uncertainties are defined: detector-related uncertainties, referred to as `experimental', and uncertainties in MC modelling of the processes, referred to as `theoretical'.
 
\subsection{Experimental uncertainties}
Multiple experimental uncertainties have been considered for this analysis, although only a small number of them have a significant impact on the results. These uncertainties are discussed below.
 
The uncertainty in the combined 2015–2018 integrated luminosity~\cite{ATLAS-CONF-2019-021} is 1.7\%, obtained using the LUCID-2 detector~\cite{Avoni:2018iuv} for the primary luminosity measurements.
Uncertainties in the reweighting procedure applied to the simulation to bring its pile-up multiplicity distribution into agreement with that in the data are also included, ranging between 0.5\% and 1\%.
 
For the leptons, uncertainties due to the measured momentum resolution and scale are taken into account~\cite{EGAM-2018-01,PERF-2015-10}. Uncertainties in the reconstruction, identification and isolation efficiency scale factors that are used to correct for the difference between the MC simulation and data are also included.
The impact of this set of uncertainties on the expected yield for the signal regions varies between 1\% and 2\%.
 
Sets of uncertainties in the jet energy scale and resolution are also included. These were derived from information taken from test-beam data, LHC collision data and simulation~\cite{JETM-2018-05,PERF-2016-04}. These uncertainties are small for all signal regions, with an impact between 0.2\% and 0.9\%.
 
Uncertainties associated with the above objects are propagated to an uncertainty in the hard term of the \met computation. Further uncertainties affecting the \met that are included are uncertainties in the offset and resolution of the soft term~\cite{ATLAS-CONF-2018-023}. 
The impact of these ranges between no impact (for the \fl, off-$Z$ region, which is insensitive to \met) and 4\% for the $4\ell$, on-$Z$, \himet region.
 
Several systematic uncertainties on the fake factors are considered. First, there is an uncertainty due to limited number of events in the single-lepton region where the fake factors are calculated. Then, there is an uncertainty in the MC modelling of the dominant background contributions in the single-lepton region (the $W+$jets and $Z+$jets contributions). The fake-lepton background estimate is compared with an estimate obtained when using fake factors that consider events with and without \bjets~\cite{FTAG-2018-01}, with the difference between the two estimates taken as an uncertainty. Finally, two uncertainties are included to address the bias caused by imposing a \met upper bound in the fake-lepton estimation regions, and by a \pt requirement on the tag jet in the fake-muon estimation regions. These uncertainties are estimated by varying the requirements on these variables upwards and downwards by 10~\GeV{}. The impact of the fake-factor uncertainties on the total background prediction ranges between 0.1\% (for the \fl regions, where there are very few fake-lepton events) and 1.6\%.
 
\subsection{Theoretical uncertainties}
 
Theoretical uncertainties affect the MC-based background estimate of the multiboson and top-quark backgrounds. The main theoretical uncertainties considered for this analysis originate from the missing higher orders in the perturbative expansion of the partonic cross-section, from PDF uncertainties and the choice of PDF, and from the uncertainty in the strong coupling constant ($\alpha_\mathrm{s}$). The analysis follows the PDF4LHC recommendations~\cite{Butterworth:2015oua} for the computation of these uncertainties. These uncertainties are uncorrelated between different background contributions. Other uncertainties such as matching and merging uncertainties, hadronisation and parton-shower uncertainties are not included as this analysis is not directly sensitive to jets.
 
For the diboson and triboson processes, the contribution of missing higher-order diagrams is estimated by observing the differences in the cross-section prediction when varying the renormalisation scale \muR and factorisation scale \muF. These scales are independently varied upwards and downwards by a factor of two~\cite{Bothmann:2016nao}, leaving out predictions where the terms are scaled in opposite directions.
This leads to a total of seven scale variations. 
The total uncertainty is taken as the envelope of all variations, picking the variation with the largest value in each individual \minv{} bin and region.
 
For the $t\bar{t}X$ ($X = W, Z, H$) contributions to the top-quark background, the uncertainty due to missing higher orders is estimated in the same way as for the diboson and triboson uncertainties, using the envelope of the seven variations of  \muR and \muF, while the PDF uncertainty is taken as the standard deviation of 100 replica variations. The impact of uncertainties in $\alpha_\text{s}$ is taken from Ref.~\cite{deFlorian:2016spz}. Predictions from alternative generators for \ttW and \ttZ processes (\sherpa~\cite{Bothmann:2019yzt} for $t\bar{t}W$, \mgamc~\cite{Alwall:2014hca} interfaced with \herwig~7~\cite{Bahr:2008pv} for $t\bar{t}Z$) are used in assessing an uncertainty due to the choice of generator; this uncertainty is found to have no impact on the final result.
For the rare top-quark processes \tZq, \ttWW and \tttt, the scale and PDF uncertainties are taken as those associated with the computed NLO cross-section values reported in Ref.~\cite{Alwall:2014hca}.
A more precise estimation of these uncertainties is considered unnecessary due to the minute contribution of these rare top-quark processes to the total background yield of this analysis.
 
The impact of the scale uncertainties on the total background estimate ranges between 5\% and 15\%; the SRs with high \met and \minv{} requirements are at the high end of this range as the impact of higher-order diagrams is especially large there. The impact of the PDF uncertainty is around 2\%--3\%, with higher values (up to 6\%) in higher \minv{} bins. The impact of the $\alpha_\mathrm{s}$ uncertainty is 1\%--2\%.
 
In general, the dominant sources of systematic uncertainty for this analysis are the theoretical uncertainties. Of these, the \muR and \muF scale uncertainty of the diboson backgrounds ($WZ$ for the \tl SRs, $ZZ$ of the \fl SRs) has the largest effect. However, many SRs are still statistically limited. This is the case for the \fl SRs (except for the \fl, on-$Z$ SRs with $\minv < 400$~\GeV); for the two \tl SRs where \lomet and $\minv > 400$~\GeV{}; and for the \tl, off-$Z$, \himet SR with $\minv > 600$~\GeV{}.

\section{Statistical analysis and results}
 
Each of the 22 SRs is treated in an individual, together with the CRs, cut-and-count experiment, and a statistical analysis is performed independently for each region.
For each of these analyses, the parameter of interest is the number of signal events in the corresponding SR: $N_{\mathrm{S}}$. The same $WZ$ and $ZZ$ control regions are used for all statistical analyses as well.
 
This analysis employs a maximum-likelihood technique, using the profile likelihood ratio (see, e.g., Ref.~\cite{kendall}) to
estimate $N_{\mathrm{S}}$
while also accounting for the various systematic uncertainties affecting the background predictions, which enter the likelihood expression as nuisance parameters (NP) $\vec{\theta}$.
The likelihood for each SR is the product of Poisson probability terms $\mathcal{P}(n|\mu)$ for the single SR and multiple control region bins.
The predicted number of events in each region is expressed in terms of the $WZ$ and $ZZ$ background normalisation factors $k_{WZ}$ and $k_{ZZ}$, the nuisance parameters and, in the case of the SR, $N_{\mathrm{S}}$.
For each systematic uncertainty, indexed by $l$, the likelihood formula is multiplied by a constraint term, which is a standard Gaussian response function
$\mathcal{G}(0|\theta_l,\sigma_l)$
with mean and standard deviation equal to the central value of the nuisance parameter and its uncertainty, respectively.
The likelihood therefore takes the form:
\begin{eqnarray*}
L & = & \mathcal{P}(N_{\text{SR}}|
N_{\mathrm{S}} + k_{WZ}N^{\text{SR}}_{WZ}(\vec{\theta}) +  k_{ZZ}N^{\text{SR}}_{ZZ}(\vec{\theta}) + \sum_{j\in \text{bg}}N_{j}^{\text{SR}}(\vec{\theta}))\\
& \times &
\prod_{i \in WZ,\, ZZ\, \text{CR bins}} \mathcal{P}(N_{i}|k_{WZ}N^{i}_{WZ}(\vec{\theta}) +  k_{ZZ}N^{i}_{ZZ}(\vec{\theta}) + \sum_{j\in \text{bg}}N_{j}^{i}(\vec{\theta}))
\; \times \; \prod_{l \in \text{NP}}\mathcal{G}(0|\theta_{l},\sigma_{l}),
\end{eqnarray*}
where the index $j$ runs over all background contributions other than $WZ$ and $ZZ$.
Many sources of uncertainty affect the predictions in both the SR and the CRs. The correlations between predictions in the different regions are accounted for by a common dependence on the associated nuisance parameter(s). The impact of these may vary between control and signal regions, particularly for the dominant systematic uncertainties (which are the diboson scale uncertainties).
 
\subsection{Likelihood fit to data}
 
Figures~\ref{fig:trex_fit_cr}--\ref{fig:trex_fit_vr} show comparisons between data and predictions for CRs and VRs after performing a likelihood fit to the CRs. The values of the normalisation factors are obtained from a binned CR-only fit (i.e.\ its likelihood formula does not include a term for any signal region) to the distributions in Figure~\ref{fig:trex_fit_cr}.
The normalisation factors are found to be $0.98 \pm 0.07$ for the $WZ$ background, consistent with the generator cross-section, and $1.05 \pm 0.09$ for the $ZZ$ background, which is consistent with previous measurements of the event yield of on-shell $ZZ$ decays~\cite{STDM-2018-30}. Furthermore, the VRs show good agreement between data and background,
indeed validating the off-shell $WZ$ modelling and the understanding of the fake-lepton background.
 
Figure~\ref{fig:fitting_postFit} shows all SRs with the observed data and post-fit background yields.
The event yields obtained by comparing the measured data with the expected background are shown, after performing the fitting procedure, in Table~\ref{tab:expected_events_postfit_3l} for the \tl signal regions, and in Table~\ref{tab:expected_events_4l} for the \fl signal regions.
 
The assumed number of signal events $N_{\mathrm{S}}$ is allowed to float freely. It is determined independently for each SR and is not required to be positive.
Given that the SR is not binned, it does not add further degrees of freedom and hence does not constrain the backgrounds further; 
the post-fit background values for every SR are therefore always the same as those of the CR-only fit, with $N_{\mathrm{S}}$ adjusted to make up for the difference between the total background and the observed data in the SR.
 
For each SR a fit finds the number of signal events ($\hat{N}_{\mathrm{S}}$) and, from the parabolic behaviour of the log-likelihood around its maximum, its associated uncertainty ($\Delta \hat{N}_\mathrm{S}$). From these values a significance can be computed, defined as $Z = \hat{N}_{\mathrm{S}}/\Delta \hat{N}_{\mathrm{S}}$ so that negative significance is associated with negative signal yields. The significances for all SRs are given in Table~\ref{tab:q0_values}. In this analysis, all SRs have a significance below $|Z| < 3$.
 
\begin{figure}[p]
\centering
\includegraphics[width=0.49\textwidth,keepaspectratio]{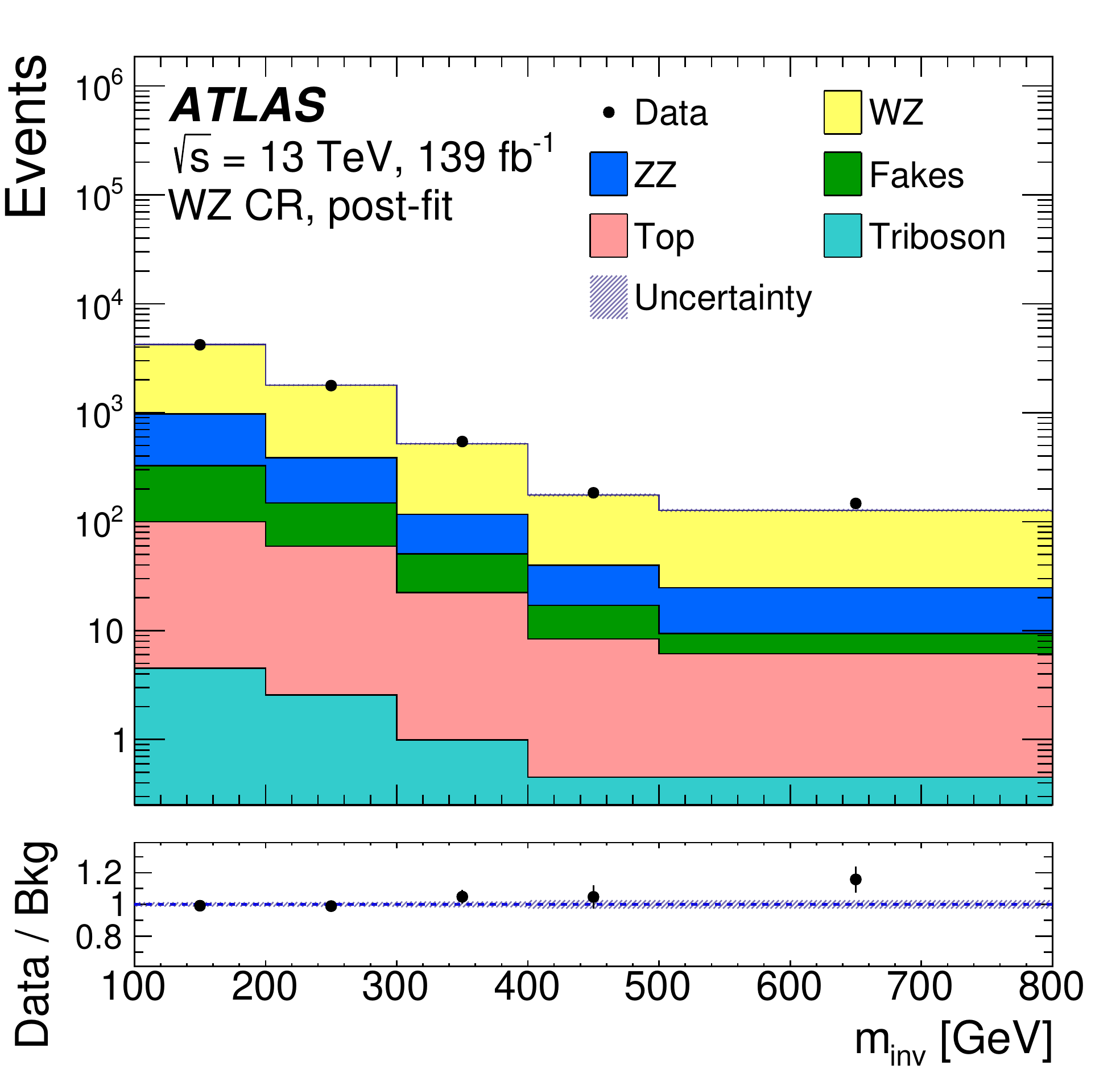}
\includegraphics[width=0.49\textwidth,keepaspectratio]{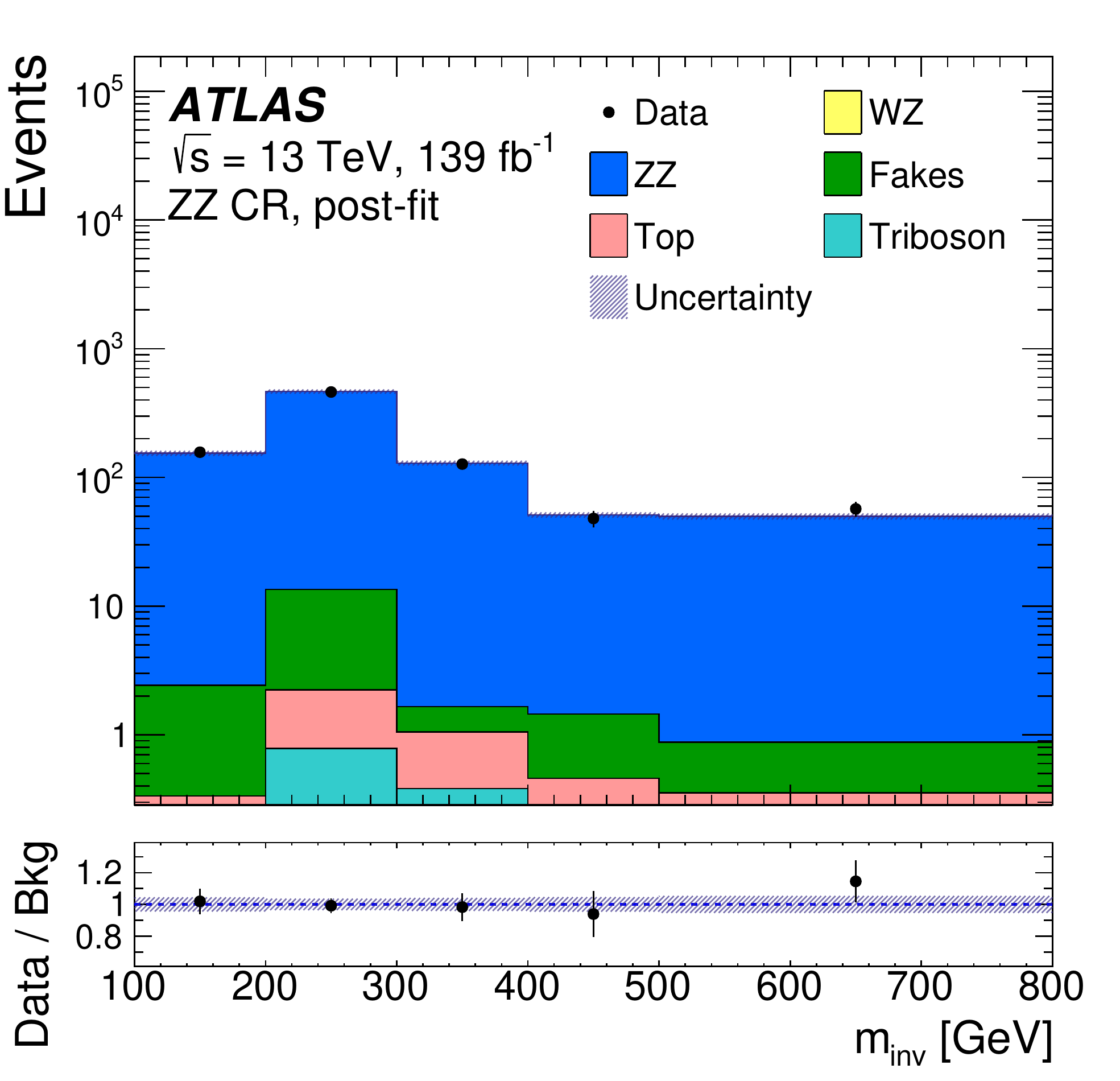}
\caption{Comparison between data and prediction for the \minv{} distribution of the $WZ$ (left) and $ZZ$ (right) control regions after the fit to the data. The rightmost bin is inclusive and contains all events with \minv{}~>~500~\GeV. `Fakes' refers to the fake-lepton background. The hatched grey area shows the combination of all uncertainties in the analysis.}
\label{fig:trex_fit_cr}
\end{figure}
 
\begin{figure}
\centering
\includegraphics[width=0.49\textwidth,keepaspectratio]{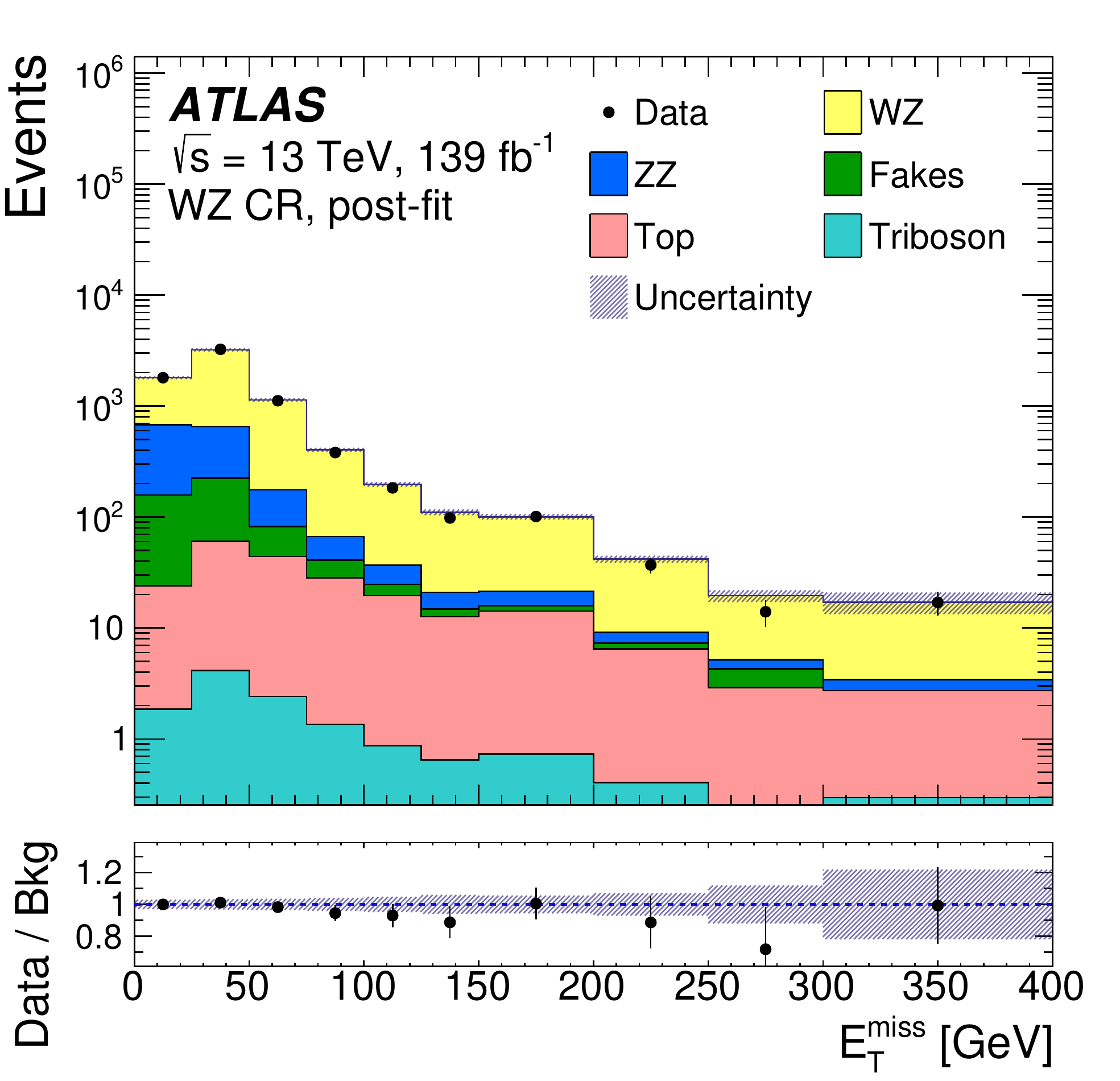}
\includegraphics[width=0.49\textwidth,keepaspectratio]{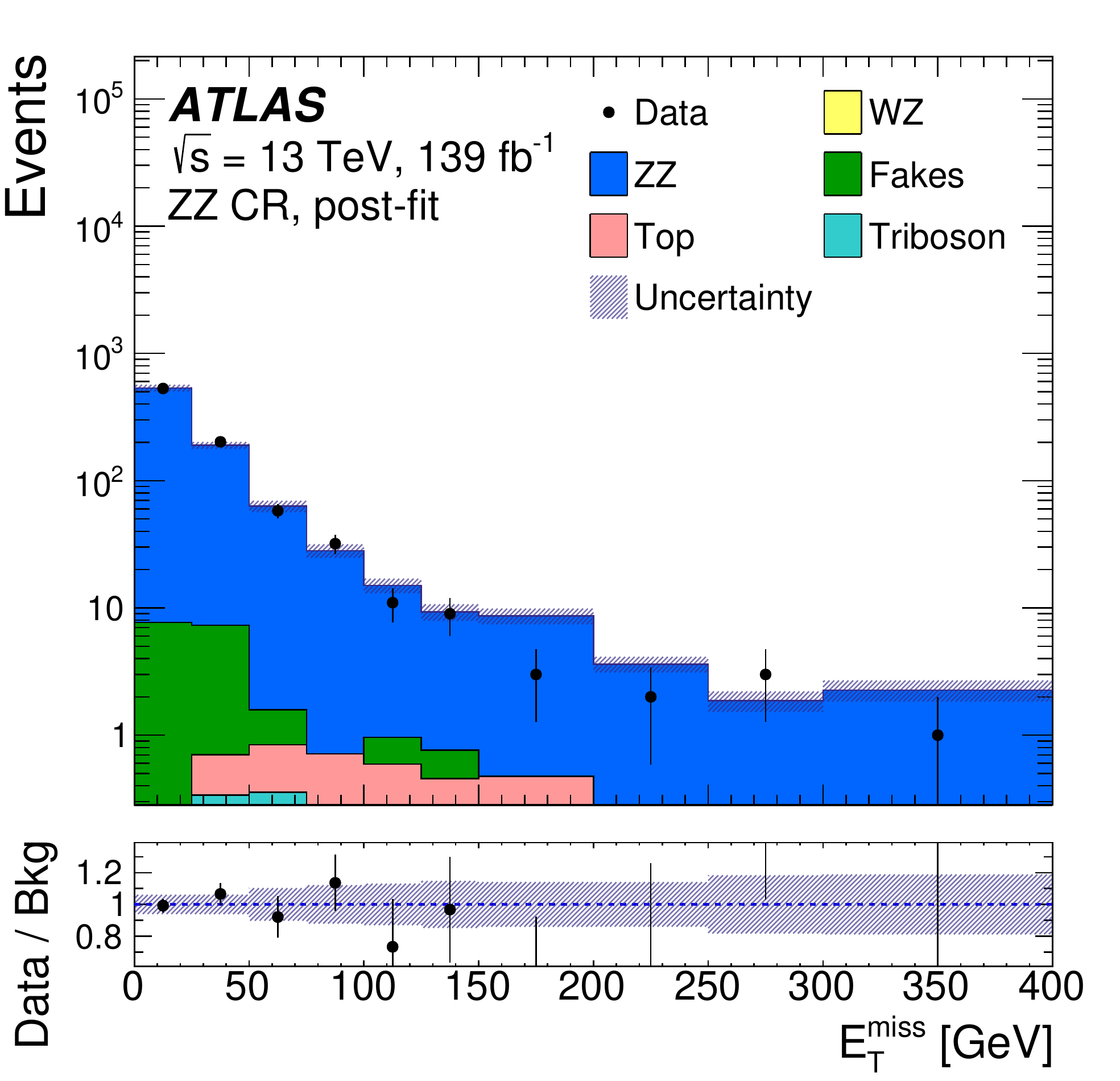}
\caption{Comparison between data and prediction for the \met distribution of the $WZ$ (left) and $ZZ$ (right) control regions after the fit to the data. The rightmost bin is inclusive and contains all events with \met~>~300 \GeV. `Fakes' refers to the fake-lepton background. The hatched grey area shows the combination of all uncertainties in the analysis.}
\end{figure}
 
\begin{figure}[p]
\centering
\includegraphics[width=0.49\textwidth,keepaspectratio]{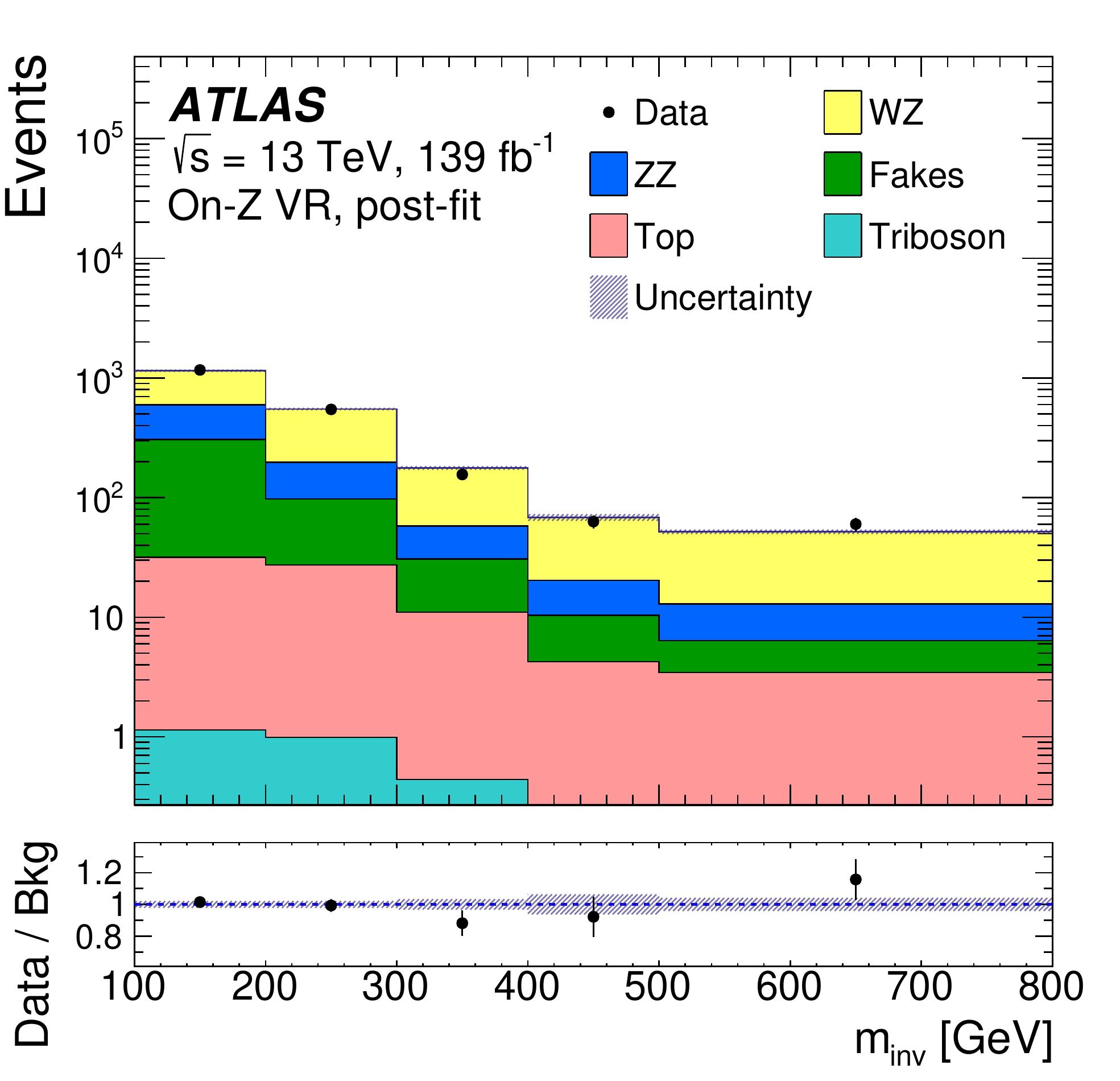}
\includegraphics[width=0.49\textwidth,keepaspectratio]{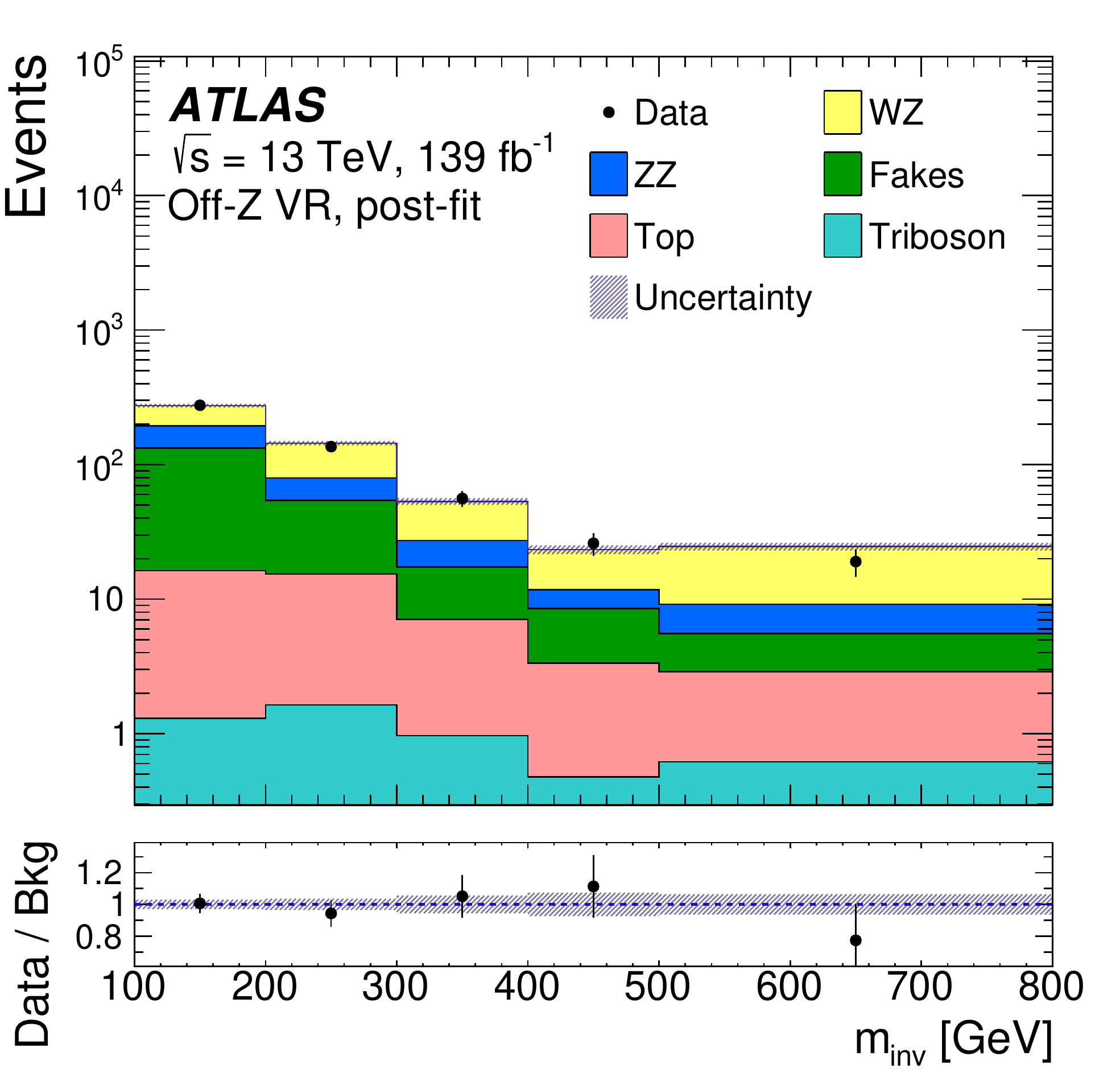}
\caption{Comparison between data and prediction for the \minv{} distribution of the on-$Z$ (left) and off-$Z$ (right) validation regions after the fit to the data. The rightmost bin is inclusive and contains all events with \minv{}~>~500~\GeV. `Fakes' refers to the fake-lepton background. The hatched grey area shows the combination of all uncertainties in the analysis.}
\label{fig:trex_fit_vr}
\end{figure}
 
\begin{figure}[hpt]
\centering
\includegraphics[width=0.8\textwidth,keepaspectratio]{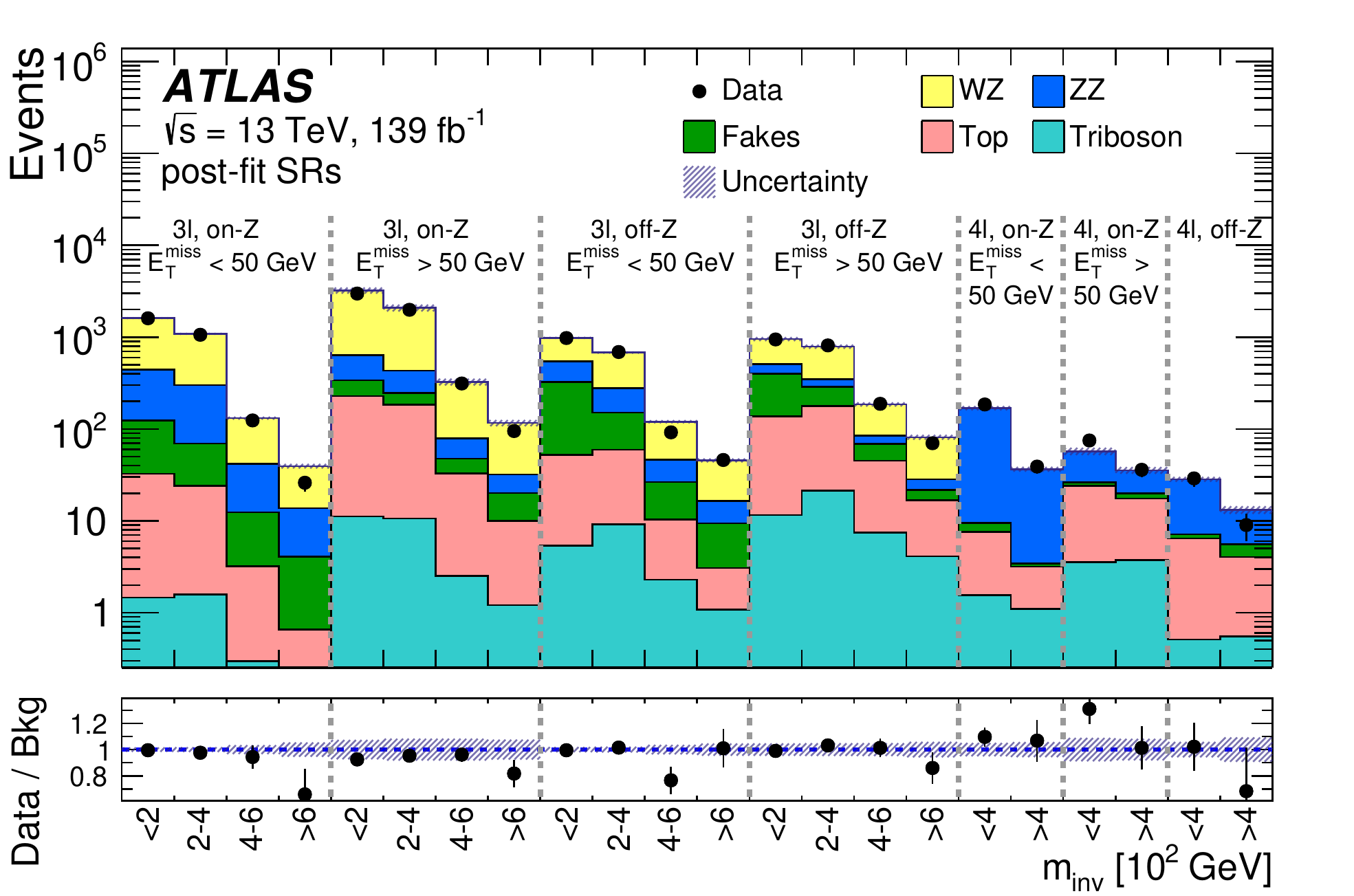}
\caption{Comparison between data and prediction in each signal region of this analysis after the profile likelihood fit has been performed. `Fakes' refers to the fake-lepton background. The hatched grey area shows the combination of all uncertainties in the analysis.}
\label{fig:fitting_postFit}
\end{figure}
 
\begin{table}[pht]
\centering
\caption{Summary of the event yields for all background contributions to the \tl signal regions after the combined likelihood fit has been performed. The observed data for each signal region are also given. `Fakes' refers to the fake-lepton background. The `Total' row gives the sum of event counts of all individual Standard Model contributions. The `Signal' row gives the remaining discrepancy between the sum of backgrounds and observed data. The total uncertainty in the event yield is given for each background contribution, for the total Standard Model prediction and for the best-fit value of the signal contribution.
It should be noted that the uncertainty in the total background prediction cannot be obtained as a sum in quadrature of the uncertainties in its contributions, due to statistical correlations.}
\resizebox{0.61\textwidth}{!}{
\begin{tabular}{lr@{ $\pm$ }lr@{ $\pm$ }lr@{ $\pm$ }lr@{ $\pm$ }l}
\toprule
\multicolumn{9}{c}{\textbf{ \textit{$3\ell$}, On-$Z$, $\met < 50$ GeV } }\\
Sample & \multicolumn{2}{c}{<200 GeV} & \multicolumn{2}{c}{200--400 GeV} & \multicolumn{2}{c}{400--600 GeV} & \multicolumn{2}{c}{>600 GeV}\\
\midrule
Top & 31.1 & 3.2 & 22.5 & 2.8 & 2.9 & 0.5 & 0.53 & 0.11\\
Triboson & 1.47 & 0.16 & 1.58 & 0.18 & 0.30 & 0.05 & 0.13 & 0.02\\
WZ & 1168 & 25 & 787 & 19 & 89.6 & 3.3 & 25.8 & 1.7\\
ZZ & 320 & 13 & 230 & 9 & 29.4 & 1.5 & 9.6 & 0.9\\
Fakes & 91 & 9 & 45 & 7 & 9.2 & 2.5 & 3.4 & 0.9\\
\hline
Total & 1612 & 26 & 1087 & 19 & 131 & 4 & 39.5 & 2.1\\
Data & \multicolumn{2}{c}{1604} & \multicolumn{2}{c}{1061} & \multicolumn{2}{c}{124} & \multicolumn{2}{c}{26}\\
\hline
Signal & -8 & 48 & -26 & 38 & -7 & 12 & -13 & 5\\
\midrule
\multicolumn{9}{c}{  }\\
\midrule
\multicolumn{9}{c}{\textbf{ \textit{$3\ell$}, On-$Z$, $\met > 50$ GeV } }\\
Sample & \multicolumn{2}{c}{<200 GeV} & \multicolumn{2}{c}{200--400 GeV} & \multicolumn{2}{c}{400--600 GeV} & \multicolumn{2}{c}{>600 GeV}\\
\midrule
Top & 217 & 19 & 174 & 18 & 30.2 & 3.4 & 8.8 & 1.2\\
Triboson & 11.2 & 1.4 & 10.6 & 1.4 & 2.5 & 0.4 & 1.21 & 0.21\\
WZ & 2589 & 216 & 1650 & 162 & 246 & 25 & 84 & 8\\
ZZ & 295 & 41 & 186 & 24 & 32 & 4 & 12.1 & 1.5\\
Fakes & 112 & 16 & 61 & 11 & 14.8 & 3.3 & 10.0 & 1.8\\
\hline
Total & 3224 & 235 & 2082 & 172 & 325 & 27 & 116 & 9\\
Data & \multicolumn{2}{c}{2982} & \multicolumn{2}{c}{1985} & \multicolumn{2}{c}{313} & \multicolumn{2}{c}{95}\\
\hline
Signal & -242 & 249 & -97 & 183 & -12 & 33 & -21 & 13\\
\midrule
\multicolumn{9}{c}{  }\\
\midrule
\multicolumn{9}{c}{\textbf{ \textit{$3\ell$}, Off-$Z$, $\met < 50$ GeV } }\\
Sample & \multicolumn{2}{c}{<200 GeV} & \multicolumn{2}{c}{200--400 GeV} & \multicolumn{2}{c}{400--600 GeV} & \multicolumn{2}{c}{>600 GeV}\\
\midrule
Top & 47 & 4 & 50 & 5 & 8.1 & 0.8 & 2.00 & 0.27\\
Triboson & 5.4 & 0.5 & 9.2 & 0.9 & 2.29 & 0.26 & 1.08 & 0.17\\
WZ & 437 & 9 & 401 & 10 & 73.7 & 3.0 & 29.1 & 1.2\\
ZZ & 222 & 11 & 126 & 7 & 19.9 & 1.5 & 7.0 & 0.9\\
Fakes & 272 & 16 & 92 & 10 & 16.1 & 3.0 & 6.3 & 1.4\\
\hline
Total & 983 & 19 & 679 & 14 & 120 & 4 & 45.5 & 2.1\\
Data & \multicolumn{2}{c}{978} & \multicolumn{2}{c}{689} & \multicolumn{2}{c}{92} & \multicolumn{2}{c}{46}\\
\hline
Signal & -5 & 37 & 10 & 30 & -28 & 10 & 0 & 7\\
\midrule
\multicolumn{9}{c}{  }\\
\midrule
\multicolumn{9}{c}{\textbf{ \textit{$3\ell$}, Off-$Z$, $\met > 50$ GeV } }\\
Sample & \multicolumn{2}{c}{<200 GeV} & \multicolumn{2}{c}{200--400 GeV} & \multicolumn{2}{c}{400--600 GeV} & \multicolumn{2}{c}{>600 GeV}\\
\midrule
Top & 126 & 11 & 156 & 14 & 38 & 4 & 12.6 & 1.7\\
Triboson & 11.6 & 1.3 & 21.4 & 2.8 & 7.5 & 1.1 & 4.1 & 0.6\\
WZ & 445 & 34 & 439 & 30 & 101 & 6 & 53 & 4\\
ZZ & 107 & 12 & 59 & 6 & 15.6 & 1.7 & 6.7 & 0.8\\
Fakes & 263 & 19 & 112 & 14 & 24 & 4 & 4.9 & 1.2\\
\hline
Total & 953 & 44 & 788 & 38 & 186 & 9 & 81 & 5\\
Data & \multicolumn{2}{c}{944} & \multicolumn{2}{c}{813} & \multicolumn{2}{c}{188} & \multicolumn{2}{c}{70}\\
\hline
Signal & -9 & 55 & 25 & 48 & 2 & 16 & -11 & 10\\
\bottomrule
\end{tabular}
}
\label{tab:expected_events_postfit_3l}
\end{table}
 
\begin{table}[ht]
\centering
\caption{Summary of the event yields for all background contributions to the \fl signal regions after the combined likelihood fit has been performed. The observed data for each signal region are also given. `Fakes' refers to the fake-lepton background. The `Total' row gives the sum of event counts of all individual Standard Model contributions. The `Signal' row gives the remaining discrepancy between the sum of backgrounds and observed data. The total uncertainty in the event yield is given for each background contribution, for the total Standard Model prediction and for the best-fit value of the signal contribution.
It should be noted that the uncertainty in the total background prediction cannot be obtained as a sum in quadrature of the uncertainties in its contributions, due to statistical correlations.}
\begin{tabular}{lr@{ $\pm$ }lr@{ $\pm$ }l}
\toprule
\multicolumn{5}{c}{\textbf{ \textit{$4\ell$}, On-$Z$, $\met < 50$ \GeV } }\\
\midrule
Sample & \multicolumn{2}{c}{0--400 \GeV} & \multicolumn{2}{c}{>400 \GeV} \\
\midrule
Top & 6.0 & 0.8 & 2.1 & 0.4 \\
Triboson & 1.56 & 0.15 & 1.10 & 0.12 \\
ZZ & 159 & 8 & 33.1 & 1.9 \\
Fakes & 1.9 & 2.1 & 0.3 & 0.7 \\
\hline
Total & 169 & 8 & 36.5 & 2.1 \\
Data & \multicolumn{2}{c}{185} & \multicolumn{2}{c}{39} \\
\hline
Signal & 16 & 16 & 2 & 7 \\
\bottomrule
\end{tabular}
\quad
\begin{tabular}{lr@{ $\pm$ }lr@{ $\pm$ }l}
\toprule
\multicolumn{5}{c}{\textbf{ \textit{$4\ell$}, On-$Z$, $\met > 50$ \GeV } }\\
\midrule
Sample & \multicolumn{2}{c}{0--400 \GeV} & \multicolumn{2}{c}{>400 \GeV} \\
\midrule
Top & 20.6 & 2.2 & 13.8 & 1.7 \\
Triboson & 3.6 & 0.5 & 3.7 & 0.6 \\
ZZ & 31 & 4 & 15.6 & 1.8 \\
Fakes & 2.2 & 1.5 & 2.5 & 1.3 \\
\hline
Total & 57 & 5 & 35.5 & 2.9 \\
Data & \multicolumn{2}{c}{75} & \multicolumn{2}{c}{36} \\
\hline
Signal & 18 & 10 & 0 & 7 \\
\bottomrule
\end{tabular}
 
\vspace{0.5cm}
 
\begin{tabular}{lr@{ $\pm$ }lr@{ $\pm$ }l}
\toprule
\multicolumn{5}{c}{\textbf{ \textit{$4\ell$}, Off-$Z$ } }\\
\midrule
Sample & \multicolumn{2}{c}{0--400 \GeV} & \multicolumn{2}{c}{>400 \GeV} \\
\midrule
Top & 5.9 & 0.7 & 3.5 & 0.5 \\
Triboson & 0.51 & 0.06 & 0.55 & 0.08 \\
ZZ & 21.3 & 0.9 & 7.6 & 0.4 \\
Fakes & 0.7 & 1.1 & 1.5 & 1.0 \\
\hline
Total & 28.4 & 1.6 & 13.2 & 1.2 \\
Data & \multicolumn{2}{c}{29} & \multicolumn{2}{c}{9} \\
\hline
Signal & 1 & 6 & -4.2 & 3.1 \\
\bottomrule
\end{tabular}
\label{tab:expected_events_4l}
\end{table}

\begin{table}[]
\centering
\caption{Local significance of the value of the parameter of interest for each signal region after performing the combined likelihood fit, defined as $Z = \hat{N}_\mathrm{S}/\Delta \hat{N}_\mathrm{S}$.}
\begin{tabular}{lcccc}
\toprule
SR   &  0--200 GeV & 200--400 GeV & 400--600 GeV & >600 GeV\\
\midrule
\textit{$3\ell$}, On-$Z$, $\met < 50$ GeV & $-0.2$ & $-0.7$ & $-0.6$ & $-2.5$\\
\textit{$3\ell$}, On-$Z$, $\met > 50$ GeV & $-1.0$ & $-0.5$ & $-0.4$ & $-1.6$\\
\textit{$3\ell$}, Off-$Z$, $\met < 50$ GeV & $-0.1$ & ~~0.3 & $-2.7$ & ~~0.1\\
\textit{$3\ell$}, Off-$Z$, $\met > 50$ GeV & $-0.2$ & ~~0.5 & ~~0.2 & $-1.2$\\
\midrule
SR   &  \multicolumn{2}{c}{0--400 GeV} & \multicolumn{2}{c}{>400 GeV}\\
\midrule
\textit{$4\ell$}, On-$Z$, $\met < 50$ GeV & \multicolumn{2}{c}{1.0} & \multicolumn{2}{c}{~~0.4}\\
\textit{$4\ell$}, On-$Z$, $\met > 50$ GeV & \multicolumn{2}{c}{1.8} & \multicolumn{2}{c}{~~0.1}\\
\textit{$4\ell$}, Off-$Z$ & \multicolumn{2}{c}{0.1} & \multicolumn{2}{c}{$-1.3$}\\
\bottomrule
\end{tabular}
\label{tab:q0_values}
\end{table}

\FloatBarrier
\subsection{Visible cross-section limits}
No significant excess was found in any of the signal regions. Limits on the number of signal events are set. The CL$_{\mathrm{S}}$ method~\cite{Read:2002hq} is used to ascertain upper limits in the signal regions. Assumptions made about the test statistic are based on the works of Wilks~\cite{Wilks:1938dza} and Wald~\cite{wald}. Specifically, it is assumed that the test statistic asymptotically approaches a $\chi^2$ distribution with one degree of freedom for a large number of events~\cite{Cowan:2010js}.
 
Expected and observed 95\% CL upper limits are presented for the visible cross-section \sigvis, which is calculated by dividing the upper limit on the total number of events ($N_{95}$) by the integrated luminosity $\mathcal{L}$ of the collected data: $\sigvis = N_{95} / \mathcal{L}$. These limits are given for all SRs in Figure \ref{fig:results_limits}. Visible cross-section limits in this figure can be reinterpreted as limits on specific physics models as long as the selection efficiency and acceptance of the model (including any uncertainties in these values) for a specific SR definition used in this analysis is known. By dividing the visible cross-section limits given here by this efficiency and acceptance, upper limits on the cross-section can be derived from this analysis.

\begin{figure}
\centering
\includegraphics[width=0.8\textwidth,keepaspectratio]{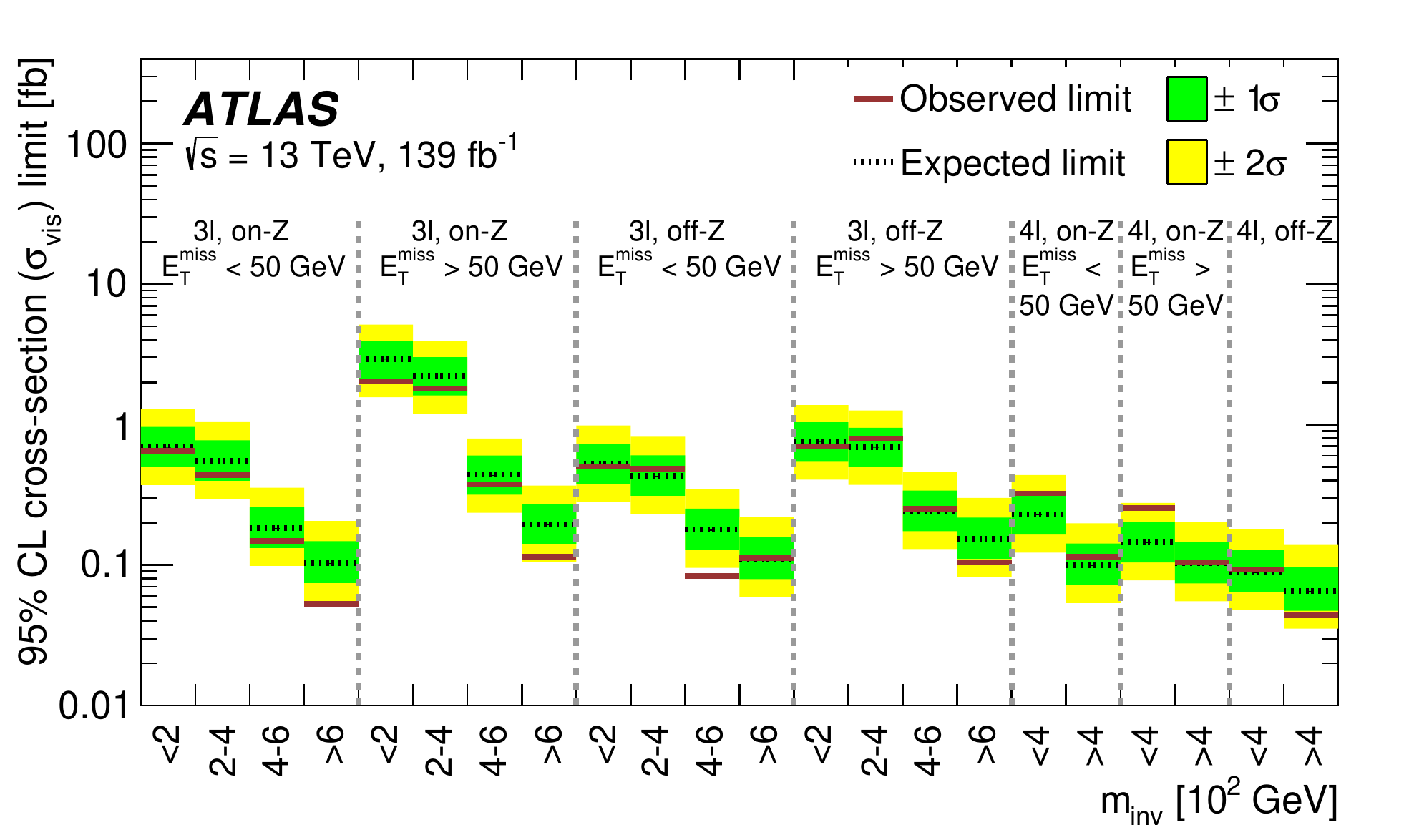}
\caption{The expected and observed 95\% CL upper limits for each signal region of this analysis expressed in terms of the visible cross-section \sigvis.}
\label{fig:results_limits}
\end{figure}
 
\FloatBarrier
 
\subsection{Model-specific limits}
 
The results are interpreted for two particular BSM models previously studied by dedicated analyses. The purpose of this interpretation is to compare the results obtained by this model-independent search with those from a dedicated search. Studied models are the Type-III seesaw model described in Ref.~\cite{t3ss} and the search for doubly charged Higgs boson production described in Ref.~\cite{EXOT-2016-07}. The chosen parameters of the models studied in this section are at parity with the referenced analyses, but only two representative particle mass hypotheses are chosen for each models: 400~\GeV{} and 700~\GeV{} for the heavy lepton in the Type-III seesaw model, and 300~\GeV{} and 500~\GeV{} for the $H^{\pm\pm}$ particles. These masses are chosen to cover a range of masses corresponding to the simulated models available and the published limits available for comparison.
 
The signal regions and background predictions remain the same as described earlier in this Letter. Using these models, all of the signal regions in these analyses are studied to find the region with the best limit-setting power, judged by the value of the expected limit. The best signal region and the corresponding expected and observed limits found by this analysis are given in Table \ref{tab:best_limits_bsm}.
To convert the limits from the model-independent analysis to limits on the cross-section of the signal model considered, only a correction for acceptance effects and selection efficiency is applied. This procedure ignores uncertainties in these quantities; however, the experimental uncertainties on the expected signal yields in the most sensitive regions are not larger than a few percent.
 
The Type-III seesaw model analysis~\cite{t3ss} presents an expected 95\% CL cross-section upper limit of $22 \substack{+8.5\\ -6.4}$~fb for $m_{L} = 400\,\GeV$ and $7.5 \substack{+3.1\\ -1.8}$~fb for $m_{L} = 700\,\GeV$ for the full Run 2 dataset with an integrated luminosity of 139~\ifb, although it only tests dilepton final states. These limits are more stringent than those derived with the analysis presented in this paper that correspond to 41 and 12~fb, respectively. The doubly-charged Higgs boson analysis~\cite{EXOT-2016-07} presents an expected 95\% CL cross-section upper limit for the four-lepton final state of $0.16 \substack{+0.14\\ -0.07}$~fb for $m_{H^{\pm\pm}} = 300\,\GeV$ and $0.14 \substack{+0.13\\ -0.07}$~fb for $m_{H^{\pm\pm}} = 500\, \GeV$. These limits are comparable to those when using the best limit of a single SR of this analysis. However, the upper limits in Ref.~\cite{EXOT-2016-07} are obtained using only 2015 and 2016 ATLAS data, corresponding to an integrated luminosity of 36.1~\ifb.
 
\begin{table}
\centering
\caption{Expected and observed cross-section exclusion limits at 95\% CL for representative mass values of the two selected models. Also the most sensitive bin, which was used to obtain these limits for each case, is listed, along with the signal acceptance times efficiency in this region (denoted by $A \times \epsilon$).}
\resizebox{\textwidth}{!}{
\begin{tabular}{l|c|cc|S|SlS}
\toprule
\textbf{Model} & \textbf{Mass [GeV]} & \textbf{Best single SR} & \textbf{\minv} & \multicolumn{1}{c|}{\textbf{$A \times \epsilon$}} & \multicolumn{2}{c}{\textbf{$\sigma^{95}_{\text{exp}}$ [fb]}}  & \multicolumn{1}{c}{\textbf{$\sigma^{95}_{\text{obs}}$ [fb]}}\\
\hline
Type-III Seesaw & 400 & \textit{$3\ell$}, Off-$Z$, $\met > 50$ \GeV{} & $>600$ \GeV{} & 0.0036 & 41  & $\substack{+17\\ -11}$ & 27 \\
& 700 & \textit{$3\ell$}, Off-$Z$, $\met > 50$ \GeV{} & $>600$ \GeV{} & 0.012 & 12 & $\substack{+5\\ -3}$ & 8.8 \\
\hline
\dch            & 300 & \textit{$4\ell$}, Off-$Z$ & $>400$ \GeV{} & 0.37 & 0.18 & $\substack{+0.08\\ -0.05}$ & 0.12 \\
& 500 & \textit{$4\ell$}, Off-$Z$ & $>400$ \GeV{} & 0.40 & 0.16 & $\substack{+0.07\\ -0.05}$ & 0.11 \\
\bottomrule
\end{tabular}
}
\label{tab:best_limits_bsm}
\end{table}
 
\FloatBarrier
\section{Conclusion}
 
In this paper, a model-independent search targeting final states with three or four light leptons is presented using 139~\ifb\ of \com $pp$ collision data collected by the ATLAS detector at the LHC from 2015 to 2018. The analysis offers a wide coverage of the \tl and \fl phase space. The measured data of \tl and \fl events is tested for potential indicators of physics beyond the Standard Model. By categorising the targeted phase space according to the number of leptons, the missing transverse momentum, the presence of a lepton pair originating from a $Z$-boson decay, and the invariant mass of the leptons in the event, a total of 22 signal regions are defined. Each signal region is analysed independently using a profile likelihood fit. Control regions were established to extract the normalisations of the primary background processes.
 
No significant deviations from the Standard Model expectation are found in the data. In the absence of a detected signal, upper limits at the 95\% CL are provided in terms of the visible cross-sections. The expected upper limits can be interpreted by dedicated analyses as long as the efficiency and acceptance of their signal model in a particular signal region is known. The analysis is interpreted using simulated signal models for heavy leptons from the Type-III seesaw mechanism~\cite{t3ss} and a doubly charged Higgs boson model~\cite{EXOT-2016-07}.

\section*{Acknowledgements}
 

We thank CERN for the very successful operation of the LHC, as well as the
support staff from our institutions without whom ATLAS could not be
operated efficiently.
 
We acknowledge the support of
ANPCyT, Argentina;
YerPhI, Armenia;
ARC, Australia;
BMWFW and FWF, Austria;
ANAS, Azerbaijan;
SSTC, Belarus;
CNPq and FAPESP, Brazil;
NSERC, NRC and CFI, Canada;
CERN;
ANID, Chile;
CAS, MOST and NSFC, China;
Minciencias, Colombia;
MEYS CR, Czech Republic;
DNRF and DNSRC, Denmark;
IN2P3-CNRS and CEA-DRF/IRFU, France;
SRNSFG, Georgia;
BMBF, HGF and MPG, Germany;
GSRI, Greece;
RGC and Hong Kong SAR, China;
ISF and Benoziyo Center, Israel;
INFN, Italy;
MEXT and JSPS, Japan;
CNRST, Morocco;
NWO, Netherlands;
RCN, Norway;
MEiN, Poland;
FCT, Portugal;
MNE/IFA, Romania;
JINR;
MES of Russia and NRC KI, Russian Federation;
MESTD, Serbia;
MSSR, Slovakia;
ARRS and MIZ\v{S}, Slovenia;
DSI/NRF, South Africa;
MICINN, Spain;
SRC and Wallenberg Foundation, Sweden;
SERI, SNSF and Cantons of Bern and Geneva, Switzerland;
MOST, Taiwan;
TAEK, Turkey;
STFC, United Kingdom;
DOE and NSF, United States of America.
In addition, individual groups and members have received support from
BCKDF, CANARIE, Compute Canada and CRC, Canada;
COST, ERC, ERDF, Horizon 2020 and Marie Sk{\l}odowska-Curie Actions, European Union;
Investissements d'Avenir Labex, Investissements d'Avenir Idex and ANR, France;
DFG and AvH Foundation, Germany;
Herakleitos, Thales and Aristeia programmes co-financed by EU-ESF and the Greek NSRF, Greece;
BSF-NSF and GIF, Israel;
Norwegian Financial Mechanism 2014-2021, Norway;
NCN and NAWA, Poland;
La Caixa Banking Foundation, CERCA Programme Generalitat de Catalunya and PROMETEO and GenT Programmes Generalitat Valenciana, Spain;
G\"{o}ran Gustafssons Stiftelse, Sweden;
The Royal Society and Leverhulme Trust, United Kingdom.
 
The crucial computing support from all WLCG partners is acknowledged gratefully, in particular from CERN, the ATLAS Tier-1 facilities at TRIUMF (Canada), NDGF (Denmark, Norway, Sweden), CC-IN2P3 (France), KIT/GridKA (Germany), INFN-CNAF (Italy), NL-T1 (Netherlands), PIC (Spain), ASGC (Taiwan), RAL (UK) and BNL (USA), the Tier-2 facilities worldwide and large non-WLCG resource providers. Major contributors of computing resources are listed in Ref.~\cite{ATL-SOFT-PUB-2021-003}.
 
 
\clearpage
 
\printbibliography
 
\clearpage 
 
\begin{flushleft}
\hypersetup{urlcolor=black}
{\Large The ATLAS Collaboration}

\bigskip

\AtlasOrcid[0000-0002-6665-4934]{G.~Aad}$^\textrm{\scriptsize 99}$,    
\AtlasOrcid[0000-0002-5888-2734]{B.~Abbott}$^\textrm{\scriptsize 125}$,    
\AtlasOrcid[0000-0002-7248-3203]{D.C.~Abbott}$^\textrm{\scriptsize 100}$,    
\AtlasOrcid[0000-0002-2788-3822]{A.~Abed~Abud}$^\textrm{\scriptsize 34}$,    
\AtlasOrcid[0000-0002-1002-1652]{K.~Abeling}$^\textrm{\scriptsize 51}$,    
\AtlasOrcid[0000-0002-2987-4006]{D.K.~Abhayasinghe}$^\textrm{\scriptsize 91}$,    
\AtlasOrcid[0000-0002-8496-9294]{S.H.~Abidi}$^\textrm{\scriptsize 27}$,    
\AtlasOrcid[0000-0002-8279-9324]{O.S.~AbouZeid}$^\textrm{\scriptsize 38}$,    
\AtlasOrcid[0000-0001-5329-6640]{H.~Abramowicz}$^\textrm{\scriptsize 158}$,    
\AtlasOrcid[0000-0002-1599-2896]{H.~Abreu}$^\textrm{\scriptsize 157}$
% ...
\bigskip
\\

% ...
$^{27}$Physics Department, Brookhaven National Laboratory, Upton NY; United States of America.\\
% ...
$^{34}$CERN, Geneva; Switzerland.\\
% ...
$^{38}$Niels Bohr Institute, University of Copenhagen, Copenhagen; Denmark.\\
% ...
$^{51}$II. Physikalisches Institut, Georg-August-Universit\"{a}t G\"ottingen, G\"ottingen; Germany.\\
% ...
$^{91}$Department of Physics, Royal Holloway University of London, Egham; United Kingdom.\\
% ...
$^{99}$CPPM, Aix-Marseille Universit\'e, CNRS/IN2P3, Marseille; France.\\
$^{100}$Department of Physics, University of Massachusetts, Amherst MA; United States of America.\\
% ...
$^{125}$Homer L. Dodge Department of Physics and Astronomy, University of Oklahoma, Norman OK; United States of America.\\
% ...
$^{157}$Department of Physics, Technion, Israel Institute of Technology, Haifa; Israel.\\
$^{158}$Raymond and Beverly Sackler School of Physics and Astronomy, Tel Aviv University, Tel Aviv; Israel.\\
% ...

\end{flushleft}


\FloatBarrier

\end{document}